\documentclass[journal]{IEEEtran}
\usepackage{url}
\usepackage{graphicx}
\usepackage{amssymb}
\usepackage{color}

\begin{document}
\title{On Decision Support for Remote Industrial Facilities using the Collaborative Engineering Framework}
\author{
\IEEEauthorblockN{Jan Olaf Blech\IEEEauthorrefmark{1}, Ian D. Peake\IEEEauthorrefmark{1}, Sudarsan SD.\IEEEauthorrefmark{7}}\\
\IEEEauthorblockA{\IEEEauthorrefmark{1} RMIT University, Melbourne, Australia}\\
\IEEEauthorblockA{\IEEEauthorrefmark{7} ABB Corporate Research, Bangalore, India}
}

\maketitle
\begin{abstract}
Means to support collaboration for remote industrial facilities such
as mining are an important topic, especially in Australia, where major
mining sites can be more than a thousand kilometers from population
centres. Software-based collaboration and maintenance solutions can
help to reduce costs associated with these remote facilities. In this
paper, we report on our collaborative engineering project providing a
decision support solution tailored for Australian needs. We present
two application examples: one related to incident handling in
industrial automation, the other one in the area of smart energy systems.
\end{abstract}


\IEEEpeerreviewmaketitle

\section{Introduction}
Means to collaborate over large distances and ways to monitor and support
remote industrial installations over large distances can improve
operation, maintenance and commissioning
costs. This is especially important in the Australian context, where
industrial installations such as mining sites, but also plants related
to power generation and agriculture, are often located far away
from major population centres. Reducing the number of on-site staff, and
the need to bring staff from the population centres into these remote
areas by facilitating remote monitoring, operation and to a lesser
extend commissioning, is the main goal of the software framework
described in this paper.

The operation of industrial systems is typically associated with
{\it events} occurring over time. We consider events of a very diverse
nature: events may comprise
automatically generated {\it alarms}, e.g., coming from a SCADA system
and associated with a malfunctioning valve. On the other hand, we can
also have manually created events such as someone pushing a help
button on a service screen. In some cases many events can occur in
a relatively short amount of time, e.g., lightning hits an
electrical substation and a large number of components
failure at literally same time. On the other hand, we are
also covering events such as a consulting request where only a
few may occur within a year.

Our collaborative engineering framework aims to support human
collaboration by processing these events with respect to
semantic carrying models and generating a response. A simple example
of a semantic carrying model would be a spatial model of a plant with
coordinates indicating the position of sensors. Once a sensor detects
something unusual, the closest staff member can be determined by
retrieving the relevant coordinates from the model and combining it
with information on staff availability. Furthermore, rules can be used
to indicate how the relevant staff members should be informed and what
information shall be displayed, e.g., on the staff member's mobile device.

The work presented in this paper summarizes the results achieved in
the collaborative engineering project which took place between 2013 and 2016
in the context of the Australia-India Centre for Automation Software
Engineering at RMIT University and the Indian ABB Corporate Research Centre.

\subsection*{Overview}
Section~\ref{sec:relwork} provides an overview of related work. Our
collaborative engineering framework is introduced in
Section~\ref{sec:ce} and the
decision support framework is described in
Section~\ref{sec:decs}. Section~\ref{sec:demo} features a description
of our demonstrator platform and a conclusion is provided in Section~\ref{sec:concl}.

\section{Related Work}
\label{sec:relwork}
Table~\ref{tab:t1} provides an overview on main directions to
support control and operation in industrial automation. We have chosen
classical SCADA-based approaches, visualization sofware-based
solutions, our collaborative engineering framework, as well as
social media-like frameworks.
The table gives
an indication of their typical application in the life cycle of a
facility, their relative strength with respect to
distributedness, their ability to take  semantic models such as ontologies into
account, the industrial automation focus, and typical interaction times. Concrete solutions are
discussed throughout this section.
We classify existing software into the following categories:
{\em Classical SCADA}, that is, supervisory control / data acquisition, is closely connected with traditional operational control, connecting with real time controller units, generating alarms in response to process status anomalies and enabling operator interaction with plant and equipment.
{\em Visualization software} refers to software which provides a mix of modeling, simulation and visualization capabilities with a focus on design.
{\em Collaborative Engineering} approaches refers to our approach of
processing events and selection and generation of information to
support human stakeholders.
{\em Social media} approaches, emerging more recently, obviously focus on people as individuals and complementing or surpassing traditional, perhaps less efficient, communication related services such as telephone directories or email.
The {\em Life cycle} attribute refers to the phase an industrial
facility is encountering, i.e. design, construction, commissioning, operations and maintenance.
The {\em Distributed} attribute refers to whether applications cater
to multiple seats, users or sites.
The {\em Semantic} attribute refers to the extent to which applications embody domain concepts such as physical models, for example through explicit ontology.
The {\em Industrial} attribute refers to the extent to which applications are targeted to industrial automation.
The {\em Timing} attribute refers to the typical event response times provided either by the applications or their users.

\begin{table*}
\centering
\begin{tabular}{l|r|r|r|r|r}
~ 				& Life cycle	& Distributed 	& Semantic  	& Industrial 	& Timing
\\
\hline
classical SCADA 		& ops 		& some		& minor 	& strong 	& subseconds--days
\\
visualization software		& all (design)	& minor 	& minor 	& medium 	& seconds--years
\\
collaborative engineering 	& all (remote ops)& strong 	& strong	& key focus 	& seconds--months
\\
social media			& all		& strong	& minor		& minor		& minutes--months
\end{tabular}
\caption{Different approaches to support control and operation}
\label{tab:t1}
\end{table*}

\paragraph{Visualization and Collaboration Software}
Relevant commercial solutions for collaboration comprise software frameworks such as Microsoft SharePoint,
Dassault Syst{\`e}me's Enovia~\cite{enovia} and
Delmia~\cite{delmia}.
While SharePoint
concentrates on providing solutions for the exchange of
documents such as texts,
the latter products are focussed on providing
geometric aspects of visual front-ends such as 3D graphics of involved plants and machinery.
In this paper we refer to functionality which relies on physical modeling
such as geometry as (partially) {\em semantic} approaches.
SharePoint also comes with social network features.
The collaboration relations between different participants
can be subjected to analysis \cite{cross} as a basis for
efficiency improvement, by identifying candidates for working
groups and for resolving resource conflicts and protecting
confidentiality.
The mentioned frameworks support
collaboration between and within organisations including the
operations control context and industrial operations sites. 
An alternative non-commercial approach to visualization for collaboration are the SAGE2~\footnote{sage2.sagecommons.org}
framework and its predecessor SAGE.
SAGE2 provides a web-based framework for hosting applications on large high resolution tiled display walls which are in principle also accessible
from multiple remote sites simultaneously including (``scaled down'') via ordinary web browsers.

Early academic approaches feature the term {\it collaborative engineering}
 in \cite{collint} and \cite{shade}. Means for sharing and
collaborative interaction on documents and other resources are
presented and most ideas  have meanwhile found entry
into commercial collaboration frameworks. Further related academic approaches comprise
examinations such as the impact of collaborative engineering on
software development (see, e.g., \cite{boochce}) and collections software challenges for the development of ultra-large scaled systems \cite{ultrace}, where collaboration is limited to the development phase.

\paragraph{Semantic Models and Ontologies}
Semantic carrying formal models are an important ingredient of our
framework. Typically on a less formal level, assigning semantics to distributed
documents and other information sources has
gained popularity in the context of the semantic web
\cite{semweb}. Semantic web-like information models can provide a basis for collaboration between
different industrial sites and facilitate the exchange of
data. Ontology-based approaches to collaborative
engineering also fall into this class and can be based on semantic web
technology such as 
\cite{sure}. The ComVantage project \cite{salmen} is developing a
mobile enterprise reference framework. It aims towards future internet
operability. To
some extent semantic annotations of data are taken into account and  applications in industrial automation exist. 
A framework for
collaborative requirements engineering, C-FaRM, is described in
\cite{c-farm}, and other background work on technologies for collaboration
can be found in \cite{cabani07,geo09}.
In addition, semantic models can be used to formalize cyberphysical infrastructures in construction, plant
automation and transport. 
Some existing real-world applications are aligned
towards a geometric representation of their components and are sometimes based on so-called 2.5 dimensional
GIS (Geographic Information System) representations (true three dimensional modelling seems far from
common practice \cite{friedman}) where the 3rd dimension $z=f(x,y)$ is represented
as a function $f$ of the 2 dimension $x$ and $y$ coordinates. However,
this
form of representation
\cite{apel} may  limit the ability to use these models as a basis for geometric, topological and information
retrieval. 
We do not limit ourself to a particular geometric representation,
coordinate or dimension system in our modelling work, but allow
different instantiations.
Related modelling approaches  include  standards such
as the Web 3D Services and the Sensor Web Enablement Architecture of
the Open Geospatial
Consortium\footnote{\url{http://www.opengeospatial.org}},
visualisation and decision support \cite{weaver}, as well as efficient data
structures for fast reasoning and decision support.
Semantic formalisms
for industry 4.0 are compared in~\cite{chihhong} and some related
guidelines to assist engineers are provided in the same work.

\paragraph{Formal Description Languages}
Different logic-based means to formalize semantic entities have been
developed. Logic formulae can incorporate spatial aspects and  are used as a
basis to formalize  our semantic models of
 industrial facilities.
The handbook of spatial logic \cite{hosl} discusses spatial logics,
related algebraic specifications, as well as applications that are 
not only limited to computer science.
Following process algebraic descriptions, another approach has been introduced in \cite{cardelli03,cardelli04}, which is used for describing spatial behavior and reasoning about
it. Process algebraic descriptions cover the
description of individual, generally asynchronously acting processes,
but 
with
distinct synchronization points as first
class citizens from a specification point of view.
In the spatio-temporal case, disjoint logical spaces are
represented in terms of expressions by bracketing structures and carry
or exchange  concurrent  process representations. Model checking
for process algebras in this context is
presented in \cite{slmc}. Complementing this, a graph-based technique for the verification
of spatial properties of finite $\pi$-calculus fragments (another
process algebra) is introduced
in \cite{gadducci}. More work on process algebraic specifications in
this context has been done in~\cite{haar}.
The establishment of specialized modal logics for spatio-temporal reasoning goes back to the
seventies. The Region Connection Calculus (RCC) \cite{Bennett}  includes
 predicates to indicate spatial separation of and topological
 relations between entities (regions). For example, RCC has predicates
indicating that regions do not share any points at all, shared points on the
boundary of regions, internal contact of regions --- where one region is included
and touches on the boundary of another from the inside,  overlap of
regions,
and inclusion. The work \cite{Bennett}  also features an overview of the relation of
these logics to various Kripke-style modal logics, reductions of
RCC-style fragments to a minimal number of topological predicates,
their relationship to interval-temporal logics as well as some
decidability results.
More results on spatial interpretations are presented in
\cite{hirschkoff} and additional decidability results can be found in~\cite{zilio}.

\paragraph{Smart Energy Systems}
One application area of our framework comprises Smart Grid
systems. Challenges regarding this topic have been outlined in
\cite{sg}. Different topics such as 
small grid sizes as a grid comprising one office building have been considered in \cite{fortiss}, needs on
security and robustness are discussed in \cite{amin}, and operational
needs for different energy sources (e.g., a focus on photo-voltaic operations
relevant for our paper is put in \cite{pv}) have been studied.
In our smart energy work, we are complementing  these research
areas. We are applying our framework to provide software support for managing
smart grids / smart energy systems, such as automatic decision support for human operators
in a control room, or in the field by using a mobile form-factor. 

\paragraph{Our Previous Work}
Previously we have published work around collaborative
engineering. Initial ideas and a first implementation are presented in
\cite{etfa} and \cite{etfa2}. The extension to smart energy systems is
described in \cite{etfa2016se,tr2017}. Our spatial
constraint solving framework BeSpaceD
is introduced and described in \cite{bespaced1,bespaced0,newoperators}. A summary of our VxLab
visualization facility can be found in \cite{vxlab,vxlab2}. The collaborative engineering platform has been used together with software defined network technology for gathering data from sensors (see \cite{sdn1,sdn2,sdn3}).
 This paper unifies the previously published work.

\section{The Collaborative Engineering Framework}
\label{sec:ce}
This section discusses our collaborative engineering framework. We
discuss the workflow and the main ingredients.

\subsection{Architectural Overview}
\begin{figure}
\centering
\includegraphics[width=.475\textwidth]{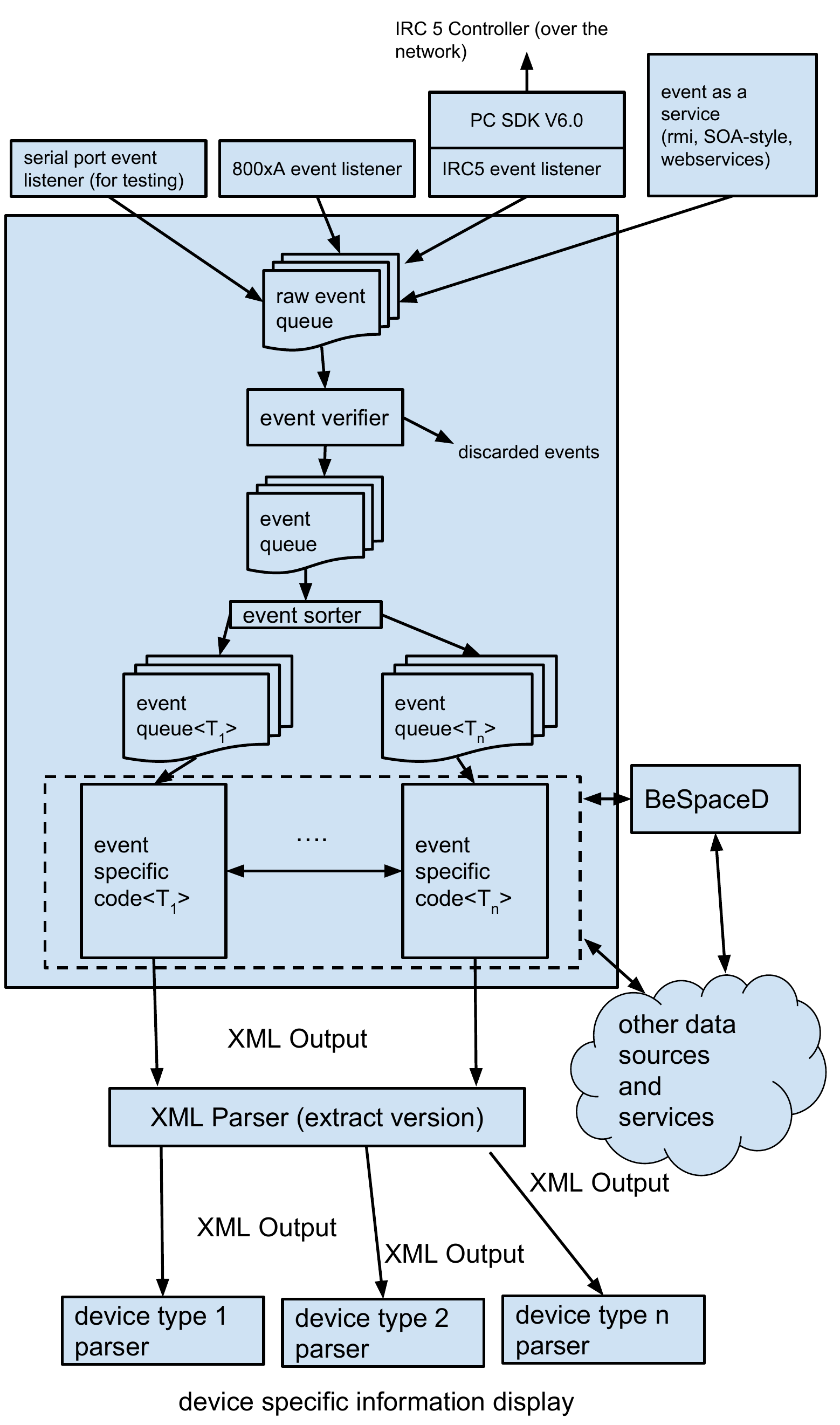}
\caption{The collaborative engineering architecture}
\label{fig:arch}
\end{figure}

Figure~\ref{fig:arch} provides an overview on our collaborative
engineering architecture. The first step involves event listeners
waiting for incoming events. Events are either triggered by various
devices such as ABB's IRC 5 robot controllers or via SOA (Service
Oriented Architecture) style
interfaces. SOA triggered events can be automatically generated or
triggered by humans. The event listeners communicate via pipes
 with the main part of the collaborative engineering framework.
In the main part of the collaborative engineering framework, events are
pulled from the pipe, sorted and queued and eventually
handed to event specific code. The event specific code uses our
BeSpaceD tool and other services to generate visualization output
which is then interpreted for various devices to trigger the display
of information.

\subsection{Visualization Information and Display}

We have designed an XML format to encode visualization and other event
related information. The outcome of the collaborative engineering
framework is encoded in this language, thereby supporting event- and
stakeholder-specific delivery of notification elements. The language
is of  scalable complexity,
ultimately up to custom applications, it can describe interactive
content and the integration of more specific collaboration tools:
Depending on an event, its location, time and the relevant stakeholders,
different notifications and representations may be needed.
The XML format provides a notation for expressing and initiating
notifications, this 
includes abstraction of stakeholder and/or location.
To provide a look-and-feel, Figure~\ref{fig:visxml} contains example
XML output for visualization.  Each command triggers the display of an
individual window on a device. The arguments allow the specification
if the device and
content. Commands can displays images with annotations {\tt
  display} such as graphical or textual overlay elements, staff
profiles, e.g., from Microsoft SharePoint and live camera views. Furthermore,
the display of google map views with coordinates and
zoom factor is encoded using the {\tt type} map and earth
commands.
\begin{figure*}
{\footnotesize
\begin{verbatim}
 <!--output>
              <command device="vxportal6" type="event" catagory="highrisk" id="1001"></command>
			  <command device="vxlab" type="display" profile="robot1"></command>
              <command device="vxportal2" type="display" profile="bob"></command>
              <command device="vxlab" type="display" profile="eric"></command>              
              <command device="vxlab" type="view" image="gridsubstation.jpg"
                    rectx="350" recty="600" rectw="120" recth="150" text="Grid Substation"
                    txtx="130" txty="90"></command>
              <command device="vxportal4" type="earth" lat="-38.1771269" long="146.3428259"
                    height="100m">
			  <command device="amritlab" type="map" lat="-38.1771269" long="146.3428259"
                    zoom="15z"></command>              
              </command>
</output-->
\end{verbatim}
}
{\footnotesize
\begin{verbatim}
<output>
           <command type="display" profile="ptz_camera3_view"></command>
		   <command type="composite_image" image="gridsubstation.jpg">
              <display type="rect" x="350" y="600" w="120" h="150"></display>
              <display type="text" text="Incident at Grid Substation" x="130" y="90" color="red"></display>
           </command>
           <command type="earth" lat="-38.1771269" long="146.3428259" height="100m"></command>
		   <command type="map" lat="-38.1771269" long="146.3428259" zoom="15z"></command>                   
</output>
\end{verbatim}
}
\caption{XML output containing visualization information (device
  independent and device specific code)}
\label{fig:visxml}
\end{figure*}

\section{Decision support}
\label{sec:decs}
This section takes a closer look at the models and rules for
decision support as well as the algorithms to support this based on
our BeSpaceD framework. It is important to note that we do not aim at
a fully automated system. Information to support decisions by humans
is provided and processed. This information may, but does not
necessarily comprise a recommendation. Ultimately decisions are taken
by humans.

\subsection{BeSpaceD overview}
BeSpaceD is our spatio-temporal modelling and reasoning framework
developed at RMIT University\footnote{\url{https://bitbucket.org/bespaced/bespaced-2016/}}. It
is a general purpose modelling and reasoning framework, thus,
applications are not limited to industrial automation or the smart-energy context.
In our work, The BeSpaceD framework is used as:
\begin{itemize}
\item a description language, for formal models of
  industrial plants,
  grids and relations between entities appearing in these models;
\item a way to reason about the formalized models by using BeSpaceD's
  libraries and functionalities.
\end{itemize}
In the following,
we describe the modelling language and the BeSpaceD-based reasoning
library functionality and provide some implementation highlights.

BeSpaceD is implemented in the Scala programming language. Scala is
bytecode compatible with Java, thus BeSpaceD's core functionality runs in a
Java environment. BeSpaceD is highly scalable; BeSpaceD-based
functionality such as services can be offered as
cloud-based services using highly scalable infrastructure, but can
also run locally on embedded devices that support a Java runtime
environment such as Raspberry Pi-based controllers. In addition to the
work described in this paper, we have successfully applied BeSpaceD to
other application areas such as
coverage analysis in the area of mobile devices~\cite{han2015} and for verification of spatio-temporal
properties of industrial robots~\cite{apscc,fesca2014,cyberbeht}.

\subsection{BeSpaceD-based Modelling}

BeSpaceD models are created using Scala case classes. 
Providing a functional abstract datatype-like feeling is a design goal
for the BeSpaceD modelling language. 
Case classes serve as abstract datatype constructors
and can be combined to create larger data structures.  
Some basic language constructs (see also \cite{2015arXiv151204656O})
are provided in Figure~\ref{fig:belang} to give a look and feel of
BeSpaceD specifications.
A key construct of our modelling framework is called an {\tt
  Invariant}, it is a basic logical entity and is
supposed to hold for a system. Despite the fact that invariants hold
for an entire system, they may contain conditional parts such as a logical formula
being part of the overall invariant
with a precondition. For example, an event occurring at a point in
space and time implies a
 state of a system. 
In our modelling methodology, constructors for basic logical operations connect invariants to form a
new invariant. Some of these basic constructors are provided in the
figure.
\begin{figure*}
{\small
\begin{verbatim}
case class OR (t1 : Invariant, t2 : Invariant) extends Invariant;
case class AND (t1 : Invariant, t2 : Invariant)  extends Invariant;
case class NOT (t : Invariant) extends Invariant;
case class IMPLIES (t1 : Invariant, t2 : Invariant)  extends Invariant;
case class TRUE() extends ATOM;
case class FALSE() extends ATOM;
case class TimePoint [T] (timepoint : T) extends ATOM; 
case class TimeInterval [T](timepoint1 : T, timepoint2 : T) extends ATOM; 
case class Event[E] (event : E) extends ATOM;
case class Owner[O] (owner : O) extends ATOM;
case class OccupyBox (x1 : Int,y1 : Int,x2 : Int,y2 : Int) extends ATOM;
case class Occupy3DBox (x1 : Int, y1: Int, z1 : Int, 
             x2 : Int, y2 : Int, z2 : Int) extends ATOM;
case class OccupyPoint (x:Int, y:Int) extends ATOM
case class Edge[N] (source : N, target : N) extends ATOM 
case class Transition[N,E] (source : N, event : E, target : N) extends
ATOM 
\end{verbatim}
}
\caption{An excerpt of logical operators for the BeSpaceD language.}
\label{fig:belang}
\end{figure*}
In the first part of the Figure, we show operators from propositional logic  (e.g., {\tt
 AND}, {\tt OR}, {\tt IMPLIES}). The second
part provides predicates for time (e.g.,{\tt TimePoint}). While the third part allows the
specification of events and ownership of logical entities,  the
fourth part provides constructs for geometry and space, such as
the {\tt OccupyBox} predicate. It refers to a rectangular two-dimensional geometric
space which is parameterized by its left lower and its right upper
corner as  x-- and y--coordinates.  The last part
provides constructs for the specification of mathematical graphs for
topologies and
state transition systems such as {\tt Edge}.

The following BeSpaceD formula expresses that the
rectangular geometric space with the corner points $(143, 4056)$ and $(1536,
2612)$ is subject to a semantic condition ``A'' between the intervals of
integer-defined time points $800$ and $950$, as well as alternatively $1000$ to $1050$.

{\small
\begin{verbatim}
  IMPLIES(AND(
        OR(TimeInterval(800,950),
           TimeInterval(1000,1050)),
               Owner("A")),
      OccupyBox(143,4056,1536,2612))
\end{verbatim}
}

The semantic condition ``A'' is a placeholder; it can be instantiated for
example by an indicator for a rain cloud or a solar panal on a
weather map.

By combining the different constructors, BeSpaceD formulae can be
constructed to formalize relevant information for our demonstrators:
such as the topologies where components influence each other, and
concrete specifications such as locations of machines,
average UV intensity in an area, specification of capacity and
location of power lines. In our framework, data can be imported while the system
is running which enables the possibility to process live streamed data
and integrate it into our decision process.

\subsection{BeSpaceD-based Reasoning}

In addition to the modelling language, we need ways to reason about the
models.
BeSpaceD provides means for the efficient analysis of BeSpaceD
formulas. The
functionality comprises ways to abstract formulas, map-reduce-like functionality, filtering and efficient
processing of information with a special focus on time and space, such as breaking geometric constraints on
areas down to geometric constraints on points. 

Various ways to import and
visualize information are supported and have been developed to program
specific needs.  BeSpaceD supports the import of information
from databases, but also the collection of information from sensors
and the conversion into BeSpaceD datastructures. 
Some tasks can be outsourced to other specialized tools  such as external
SMT solvers (e.g., we have a connection to z3 \cite{z3}) for external
processing of some information. In the SMT case,  resolving geometric
constraints such as deciding whether there is an overlapping of
different areas in time and space is a task.

\subsection{An Example Use Case}
\label{sec:exuc}

\begin{figure*}
\centering
\includegraphics[width=0.8\textwidth]{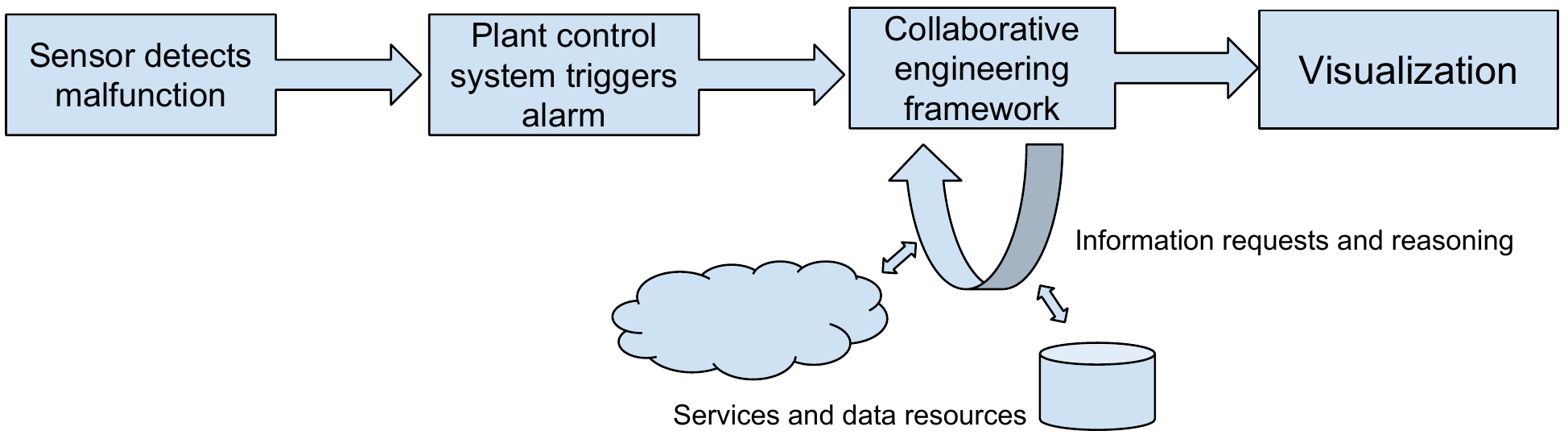} 
\caption{Collaborative engineering workflow}
\label{fig:frameworkwf}
\end{figure*}

To illustrate the workflow in our framework we are using an example use-case. 
Figure~\ref{fig:frameworkwf} shows our framework responding to an
event, an alarm triggered by a machine malfunction (see also \cite{etfa}).

A simple example is provided below (cf. \cite{etfa}):
\begin{enumerate}
\item
A sensor sends a signal to a conroller, the controller informs the
plant control system via an internal network, thereby generating an
alarm which is observed by collaborative engineering. Alternatively
the sensor controller directly communicates
with a cloud instantiation of the collaborative engineering framework, thereby indicating a machine malfunction in a remote
plant. Based on the provided information  we check the confidence by
investigating historical data from the sensor and by using data
collected from nearby sensors. 

\item
Our goal is to provide information to staff,
stakeholders, experts,
and/or engineers to facilitate the handling and collaboration among
the stakeholders. Information is preferably provided in a concise
visual way. For example, we can face situations where many alarms arrive
in a short time such as a lightning strike to a major electrical
component in a plant and information for display has to be filtered
accordingly so that humans are not overburdened with information. 
The information provisioning is customized: different human
stakeholders can be chosen by capabilities, availability or
proximity, in case a physical investigation is necessary or remote
support is sufficient. By using our BeSpaceD based reasoning, the collaboration platform can automatically
match experts to the situation and offer resource conflict resolution
while alerting them. The BeSpaceD-based reasoning can
take additional information into account, e.g., information provided
in databases including semantic models of plants and processes and
real-time information from streaming sources. Semantic models of
plants provide  mathematical logical descriptions of aspects of the
plant such as its geometric layout, cable and pipe interconnections,
information how this changes over time,
information on states of machinery (e.g., on,off, scheduled maintenance), possible dangers, possible interactions, physical
locations, and possible effects on the surrounding area. Semantic
models can also include timed information indicating how these aspects
change over time.

\item
In order to display information, we
generate incident-relevant information and encode in XML what shall be displayed to humans. We
use XML commands that are interpreted and trigger changes to
display information on mobile, devices, normal workstations or large
scale visualization facilities.
\end{enumerate}
For this use case, typical information displayed comprises profiles and other data
stored in Microsoft SharePoint, camera views,  as well as maps
(including annotations). Our Microsoft SharePoint-based data is triggered in browser
windows opened by our BeSpaceD framework. It can than
be handled interactively, for example, BeSpaceD may trigger the display of an editable collaborative document on
multiple devices in different sites.

\section{Demonstrator Platforms}
\label{sec:demo}
This section describes some of our demonstrator work.

\subsection{Demonstrator Platform}

RMIT set up the VxLab~\cite{vxlab2}. In the context of collaborative engineering
VxLab serves as a space for demonstrators.
The VxLab is a distributed lab 
developed to 
enable training,
combining high resolution visualization, industrial automation facilities and cloud-based compute servers
for analytics, visualization and simulation, connected via dedicated private networks.
VxLab
supports the collaborative engineering project.

The goals of the VxLab approach are:
\begin{itemize}
\item provision of a ``sandbox'' capability enabling rapid prototyping while mitigating security risks;
\item network interconnectivity of complementary infrastructure such as industry labs, cloud and visualization facilities not commonly combined in research or industry practice;
\item access to remote or dangerous labs without requiring physical travel, risk assessments, safety inductions or personal protective equipment;
\item for university-industry collaborations, applications in training and education, in particular through projects giving students access to current and next-generation capabilities.
\end{itemize}

VxLab design concerns include: 
(i) Safety and security---which are typically connected with legislation, industry standards and mission critical requirements, and are directly applicable where VxLab extends into industrial applications: e.g. robot labs, or data collection from SCADA servers. Security has implied closed networks and firewalls which has a major impact on capability and usability.
(ii) Latency---inherent in collaboration over distance, it is a high priority to 
minimize or mitigate latency to enable (and test) real-time machine-to-machine interoperation, human-human and human-machine interaction.
Use of pre-cached content such as models and video sent prior to a collaboration task has been demonstrated in proof of concept form.
Off-the-shelf solutions such as cloud-based video conference
tools have been preferred. Note, however, that in this work, we do not aim at hard
real-time responses. Especially, we do not want to eliminate a
human-in-the-loop; the goal is rather to support humans in their work.
(iii) Bandwith, connectivity between cloud infrastructure and external
labs and devices.
(iv) High quality visualization---to support monitoring and diagnosis of equipment such as robots, video 
quality should be maximized---in phase 1 a minimum of 1080p HD video quality was 
specified.
(v) Usability---as skilled users migrate through different roles and physical locations in distributed projects,
a usable and consistent virtual work environment must be preserved across locations.

The experience-oriented approach of VxLab has driven the exploration of use and adaptation of a mix of points on an architecture spectrum including (i) a large number of desktop and service-based local network applications, (ii) open source service- or web-based technologies and (iii) cloud technologies such as IBM Bluemix or services such as those focused on open source developer support. This has enabled the experience of different architectural approaches and consequent tradeoffs among the above concerns.

\begin{figure*}
\centering
\includegraphics[width=1.57\columnwidth]{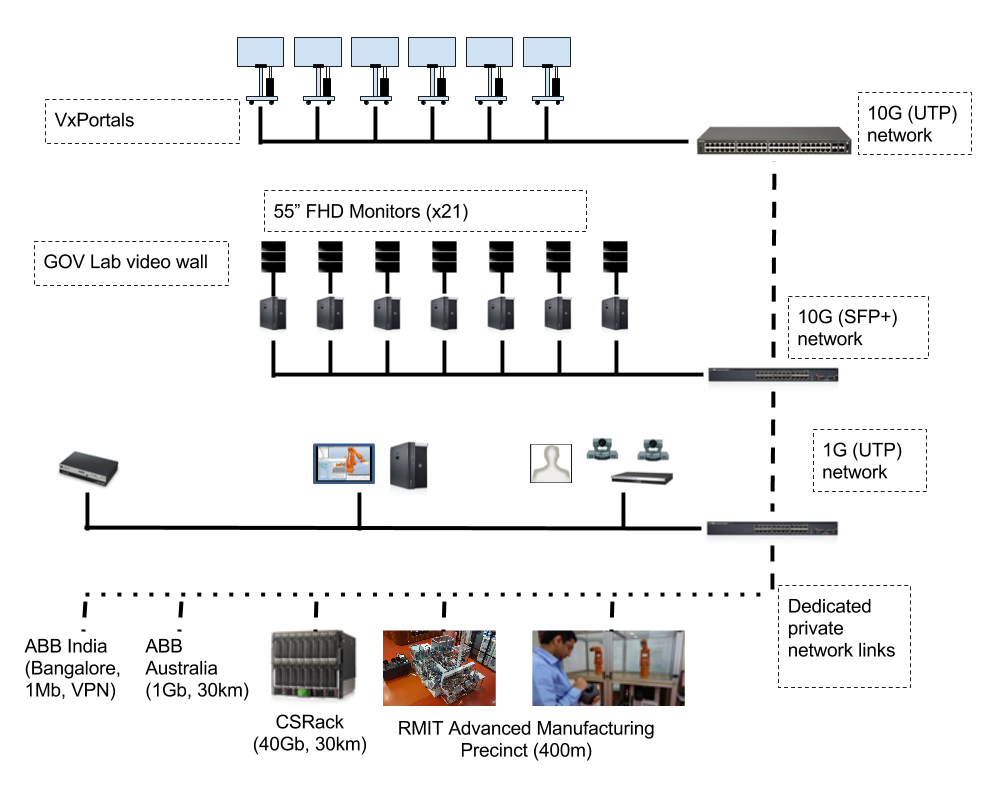}
\caption{VxLab Architecture}
\label{fig:vxlab-arch}
\end{figure*}
VxLab's network architecture is shown in Figure~\ref{fig:vxlab-arch}:

%
(i) Within VxLab, the Global Operations Visualization (GOV) Lab 
provides a high resolution 8m x 2m video display wall and PC cluster supporting multiple local users simultaneously displaying and interacting with standard applications and internet services.
GOV Lab is the primary visualization facility for the collaborative engineering project.
Figure~\ref{fig:screenshot2} shows GOV Lab.
The video wall is based on the web-based SAGE2 framework\footnote{\url{sage2.sagecommons.org}} which is scalable with respect to display wall size, number of applications and services, and number of users interacting with the display.
The SAGE2 framework
provides a method for deployment and co-ordination of active tiled applications
and supports flexible size, layout and compositing of displayed applications.
Application windows can be arbitrarily rearranged and sized across the display wall (irrespective of physical monitor boundaries) and display elements can incorporate transparency.
The display wall is driven by PCs running Linux supported by a dedicated 10Gbps switch. Each display column is driven by a separate display PC running Linux.
Applications are distributed across external clients, which provide input, data or video sources, the SAGE2 server itself, and display tiles.
The SAGE2 NodeJS web server is responsible for tracking and managing application layout, routing video and input/state sharing events via the websockets HTTP extension for real time TCP-based communication in a web context.
For rendering, display tiles run a customized web browser.
SAGE2-compliant applications send display output via the server to display tiles. Since applications are HTML5 embedded in displays it is possible to rapidly prototype and deploy rich tiled applications including 3D animations using extensions such as WebGL.
In VxLab a SAGE2 virtual desktop client is used to connect to workstations running the VNC protocol, thus existing applications for development, simulation and data analytics can be combined with video feeds for live monitoring of connected facilities. The use of standard protocols for desktop sharing enables the use of virtualized host desktops running anywhere in VxLab, subject to bandwidth/latency limitations. Beyond adaptation of existing applications,
we have demonstrated that SAGE2 enables
combinations of either or both of
(a) rich 2D or 3D applications such as interactive maps or 3D model viewers
(b) with multiple users in multiple locations separated by thousands of kilometres and high latency.
(ii) A 1Gbps local area network switch connects the video wall to other local VxLab equipment for example video conference and hardware-in-the-loop simulator, as well as a collection of services
such as SharePoint and ABB's RobotStudio IDE.
Cloud-based video conference software integrates with traditional H323 video conference hardware and enables basic sharing of the video wall content via steerable front and rear cameras.
(iii) The VxLab is connected to a set of industry and industrial labs.
For example a mini-factory which provides sensor data for the collaborative engineering framework.
A wireless internet gateway connects to a set of single board computer-based controllers connected via custom I/O boards to factory components such as pick and place units and conveyors. This enables experimental cloud-based analytics, control and visualization of plant status, including to mobile devices. Two ABB IRB120 industrial robot arms are also used to feed events into the collaborative engineering framework via IRC5 controllers. The robots have been used to demonstrate remote simulation, configuration and operation including service-based synchronisation of physical and simulated equipment. A ROS-based collaborative robot enables high bandwidth instrumentation of robot joint encoders and 3D visualization of robot configuration in real time via ROS rviz.
(iv) An additional visualization facility for collaborative engineering is provided by a cluster of 6 high resolution virtual experience portals (VxPortals)~\cite{salento}, each consisting of a 4K 60hz display, a workstation and a depth-based sensor for tracking a small number of human users. The VxPortals are connected to VxLab and each other via a 10Gbps network. The VxPortals have been used for example with a highly customized version of SAGE2 to display immersive 3D applications stretching across multiple VxPortals based on parallax via head tracking---thus the VxPortal functions as a ``magic window'' (or virtual/mixed reality with headset for a single user) into a virtual 3D rendered environment.
(v) The Cyber-Physical Simulation Rack (CSRack) provides a privately hosted publicly accessible cloud capability to support modelling, simulation and collaboration services.
CSRack consists of a 40 node cluster of blade servers with 100G of RAM and solid state disk, in a data centre connected to VxLab via shared 4x10Gbps link. Experimental OpenStack configurations provide some consistency of user experience with the Australian Nectar cloud infrastructure\footnote{\url{nectar.org.au}}, enabling for example prototyping the integration of computer vision into automation applications and the use of hosted message queues either in CSRack or in Nectar for cloud-based distributed sensor monitoring or analytics.
BeSpaceD can be deployed in Nectar or CSRack and used as a service by monitoring equipment such as controllers.
A future possibility is to explore the use of CSRack as a ``cloudlet,'' (or fog) a locally-connected (low-latency high-bandwidth) proxy to Nectar.
(vi) Advanced user interface devices such as depth sensor-based controllers, and virtual reality (VR) and augmented reality (AR) headsets are being used in the lab. VR and AR devices are currently used for gaming research with efforts underway to explore multi-disciplinary crossover into automation.


\subsection{The SmartSpace Demonstrator}
\label{sec:demosmart}

We have built a demonstrator for smart-energy decision support based on the ingredients described in the
previous sections. 
Decision support uses the collaborative engineering framework and is triggered by regularly
recurring feeds of live weather data: we have implemented a connection to the
Australian Bureau of Meteorology. BeSpaceD is used as a format to
represent the weather data and the conversion happens in real-time
Furthermore, the BeSpaceD language is used to describe
rules and models for our smart-grid system. Rules and models can be changed,
added and removed. Rules and models integrate into the Scala-based
environment. 
We have experimented with a variety of rules, for example we have used
the one provided below \cite{etfa2016se}: \\
{\small
$t_1 \le time \le t_2$ $\wedge$ \\
 {\tt cloud coverage filtered by area} $_1$ $\ge$ {\tt threshold} \\
 $\wedge$ 
 ... $\wedge$  \\ {\tt cloud coverage filtered by area}$_n$ $\ge$ {\tt
   threshold} \\
$\longrightarrow$ \\
{\tt critical solar energy level} 
}


\begin{figure*}
\centering
\includegraphics[width=.99\textwidth]{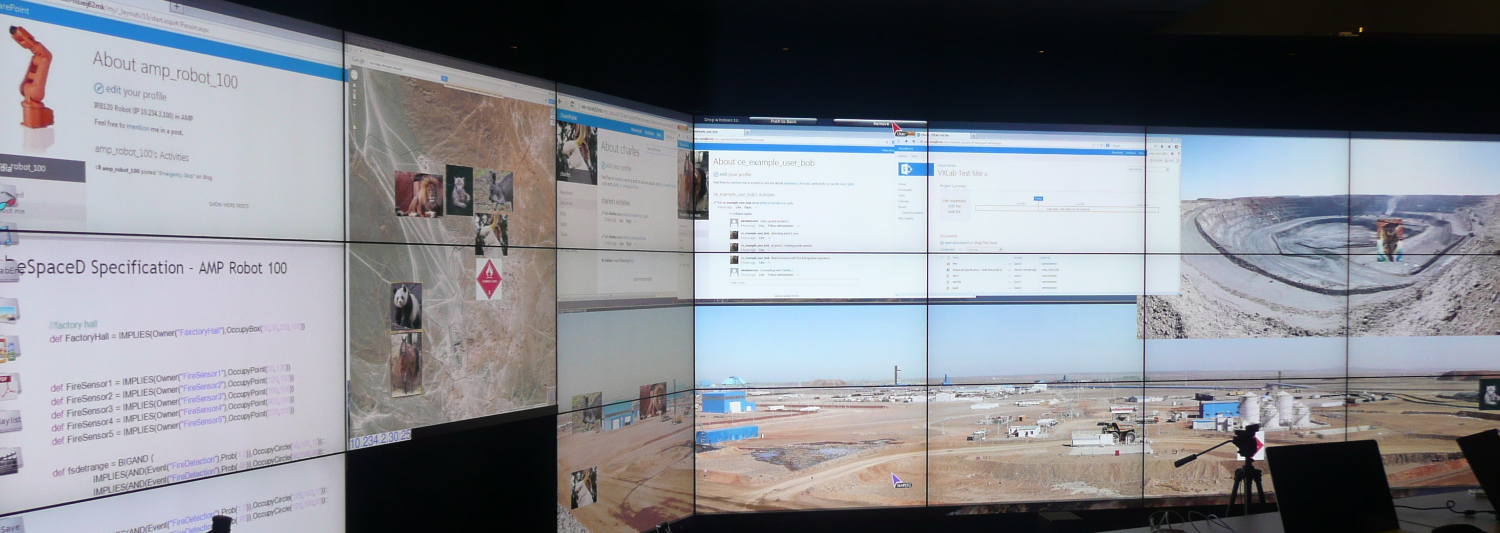}
\caption{Triggered windows in VxLab}
\label{fig:screenshot2}
\end{figure*}

The rule is provided using assumption logic on a very abstract
specification level. The form schema of the rule follows the pattern: a condition implies / triggers a
reaction. In this case,
 $t_1$ and $t_2$ are timepoints (the first line specifies a time
 interval), {\tt area}$_1$ .. {\tt area}$_n$ are spatial / geometric
areas, e.g., on a map. Furthermore, the critical solar energy level triggers a reaction.
Using this rule, based on rain-cloud coverage in some areas
stakeholders can be informed. 

The implementation of the rules is done in Scala using the constructs
provided in previous sections, for example, one needs to specify how
the filtering of an area is done by instantiating the BeSpaceD code. The shown
rule is relatively simple, in addition to the rules one also needs to specify the
triggered reactions: the XML code that comprises visualization
information and is then interpreted and
triggering a visualization.

Figure~\ref{fig:screenshot2} provides an example for a triggered
reaction in the collaborative engineering framework. Multiple windows
with profiles of staff and a information for a machine is shown. In
addition, the triggered visual information features an annotated map.
\begin{figure*}
\centering
\includegraphics[width=.7\textwidth]{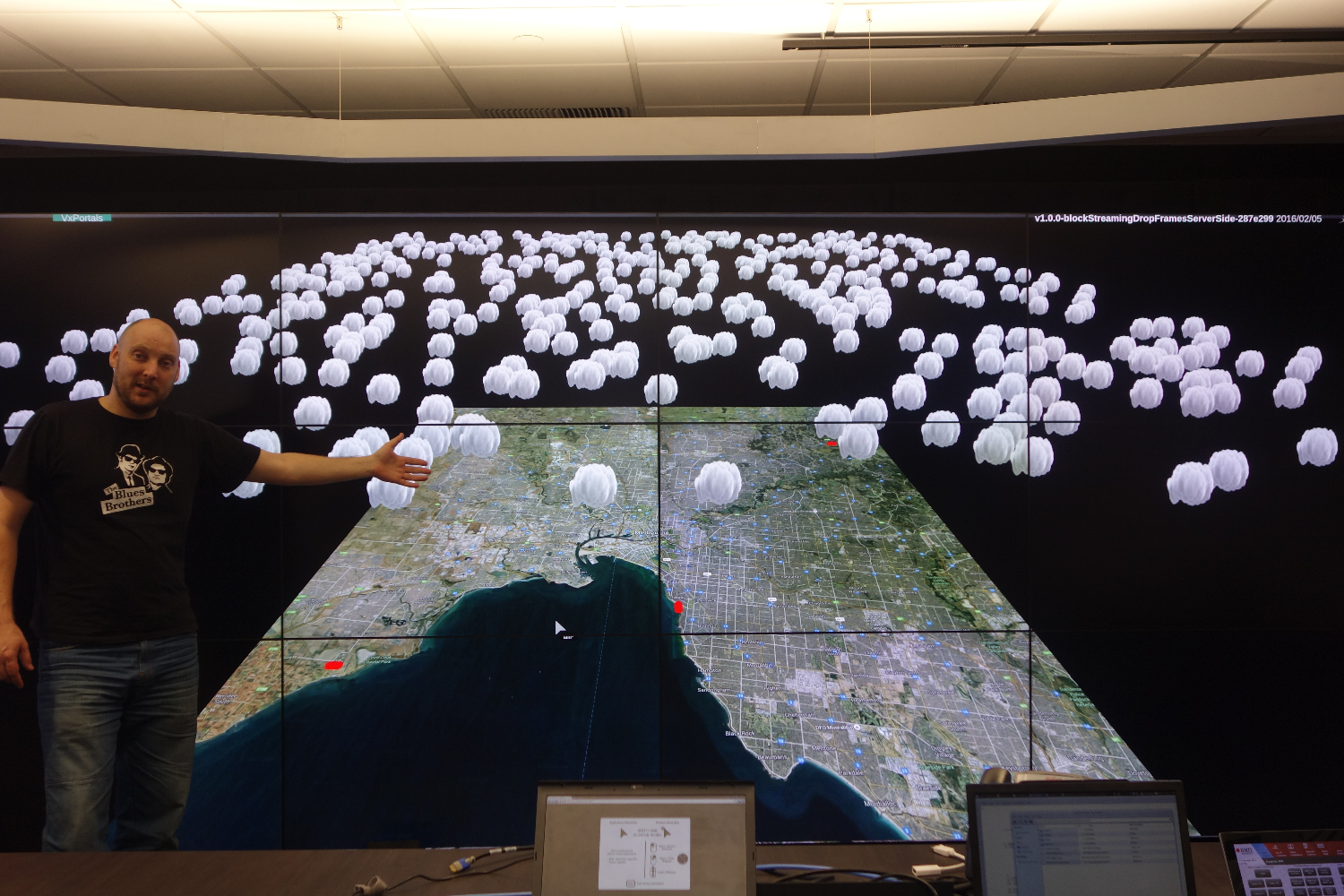}
\caption{Visualization of BeSpaceD models for weather cloud coverage}
\label{fig:ed}
\end{figure*}
Figure~\ref{fig:ed} shows another way of visualizing weather data. 
We have implemented this fronted for the SmartSpace
demonstrator. The visualization takes descriptions provided in our
BeSpaceD language as input. Here,  information on rain-cloud coverage, their position in space and
locations of power plants are contained. The generated 3D view can be
animated by adding time information to the rain-cloud positions in the
underlying BeSpaceD datastructures. A head tracking system based on
Microsoft Kinect technology can be used to move the view, so that the
perspective changes with the position of the observer.

\section{Conclusion and Future Work}
\label{sec:concl}
We presented our collaborative engineering framework and the BeSpaceD
tool-set as a main ingredient. Collaborative engineering supports
remote industrial operations. Specifically when industrial sites are
located far away from major population centres, software-based collaboration and maintenance solutions can
help to reduce costs associated with these remote operations.  We presented
our demonstrator platform infrastructure and provided examples in decision support for
industrial automation and smart-energy systems.
Future work comprises the support of additional monitoring
functionality. Monitoring of machines through cloud-based services and
connecting this to the collaborative engineering framework is an
ongoing topic. Monitors are able to detect abnormalies in
communication and timing behavior that can indicate malware and other security
violations.

\subsection*{Acknowledgement}

The authors would like to thank  Lasith Fernando, Edward Watkins, Keith Foster, Abhilash
G, and Yvette Wouters for their help.
\bibliographystyle{plain}

\end{document}